\documentclass[a4paper,fleqn,usenatbib]{mnras}
\usepackage[T1]{fontenc}
\usepackage[T2A]{fontenc}
\usepackage[utf8]{inputenc}
\usepackage[english]{babel}
\usepackage{ae,aecompl}
\usepackage{graphicx}
\usepackage{amsmath}
\usepackage{amssymb}
\usepackage{array,tabularx}
\usepackage{ulem}
\usepackage{threeparttable}
\usepackage{enumitem}
\usepackage{xcolor,cancel}

\makeatletter
\let\TPT@hookin\@gobble
\let\TPT@hookarg\@gobble
\makeatother

\title[SNe Ia: evolution]{The evolution of relative frequencies of ONe and CO SNe Ia}

\author[A. I. Bogomazov \& Tutukov]{A. I. Bogomazov$^1$\thanks{E-mail:a78b@yandex.ru (AIB)}, A. V. Tutukov$^2$\thanks{E-mail:atutukov@inasan.ru (AVT)}
\\
$^1$Lomonosov Moscow State University, Sternberg Astronomical Institute, 119234, Universitetkij prospect, 13, Moscow, Russia\\
$^2$Institute of Astronomy, Russian Academy of Sciences, 119017, Pyatnitskaya st., 48, Moscow, Russia\\
}

\date{Accepted . Received ; in original form }

\pubyear{2022}

\begin{document}
\label{firstpage}
\pagerange{\pageref{firstpage}--\pageref{lastpage}}
\maketitle

\begin{abstract}
In this population synthesis work we study a variety of possible origin channels of supernovae type Ia (SNe Ia) Among them mergers of carbon-oxygen (CO) and oxygen-neon (ONe) white dwarfs (WDs) under the influence of gravitational waves are considered as the primary channel of SNe Ia formation. We estimated frequencies of mergers of WDs with different chemical compositions and distributions of masses of merging WDs. We computed the dependence of the ratio of merger frequencies of ONe and CO WDs as primaries in corresponding binaries on time. The scatter of masses of considered sources (up to the factor $1.5-2$) of SNe Ia is important and should be carefully studied with other sophisticated methods from theoretical point of view. Our ``game of parameters'' potentially explains the increased dimming of SNe Ia in the redshift range $z\approx 0.5-1$ by the changes in the ratio of ONe and CO WDs, i.e., to describe the observed accelerated expansion of the Universe in terms of the evolution of properties of SNe Ia instead of cosmological explanations. This example shows the extreme importance of theoretical studies of problems concerning SNe Ia, because evolutionary scenario and parameter games in nature potentially lead to confusions in their empirical standardization and, therefore, they can influence on cosmological conclusions.
\end{abstract}


\section{Introduction}
\label{sec1}

Supernovae stars were divided into two main subclasses \citep{minkowski1941}: without hydrogen spectral lines (type I) and with such lines (type II). The type I is a heterogeneous group \citep{bertola1964} consisting of at least two subgroups \citep{elias1985}, the type Ia demonstrates the presence of an absorption line Si II 6150 \r{A}, the type Ib does not have this spectral line \citep{porter1987}.

According to phenomenological studies the maximum stellar magnitude of SNe I (Ia) can be computed using the observed slope of the light curve \citep{pskovskii1984,phillips1993}. This fact made it possible to use SNe Ia as standardiseable  candles for cosmological studies. Based on this method, the accelerated expansion of the Universe was introduced \citep{riess1998,perlmutter1999}, cosmological applications of SNe Ia currently play very important role in modern physics and astrophysics \citep{howell2011}.

Early models of supernovae type I (this class was not divided into subclasses that time) considered a thermonuclear explosion \citep{hoyle1960,arnett1968} of a carbon-oxygen (CO) white dwarf (WD) with super-Chandrasekhar mass in a semi-detached binary system \citep{whelan1973} and the merger of binary CO WDs (with the super-Chandrasekhar sum of masses) under the influence of gravitational wave radiation \citep{tutukov1981,iben1984b,webbink1984}.

Sub-Chandrasekhar double-detonation models were suggested later \citep{taam1980,livne1990}. The explosion of the helium layer with mass up to $0.1 M_\odot$ can trigger the detonation of CO WD with mass less than the maximum limit \citep{nomoto1982,iben1991,tutukov1992a,woosley1994}. The minimum mass of this layer (required for the detonation of CO mantle of the dwarf) changes from $\approx 0.08 M_\odot$ (for the solar mass dwarf) to $\approx 0.01 M_\odot$ (for the dwarf with mass $\approx 1.3 M_\odot$) \citep{tutukov1992b,fink2010}. Currently the double detonation of nuclear fuel as the cause of explosions in substantial partof SNe Ia is widely considered in the literature \citep{li2021,magee2021,gronow2021,sanders2021,bravo2022,collins2022}.

Another possibility for the detonation of dwarf's nuclear fuel with sub-Chandrasekhar mass is the edge explosion of the dwarf that is heated by the active accretion of helium, carbon and oxygen, and by tides \citep{iben1998,kulikov2019,tanikawa2019}. As the result of such explosion the lost envelope can take an asymmetric crescent shape that sometimes can be observed \citep{alsaberi2019,guest2022}. A possibility of the nuclear fusion in the mantle requires future 3D numerical studies. In particular, it is essential to know the minimum mass of a CO or ONe (oxygen-neon) WD that can go through the double detonation of nuclear fuel. A numerical study showed that if the mass of the WD is $\lesssim 1 M_\odot$ the double detonation cannot start \citep{fenn2016}. However, the minimum mass of the exploding dwarf in the double detonation model can be a free parameter of calculations and the model can give a reasonable coincidence with observations in a certain range of masses of exploding WDs (e.g., \citealp{polin2019}, $0.85-1.2M_\odot$ for the thin helium shell model).

At the maximum SNe Ia can be dimmer by several tenths of stellar magnitude in passive galaxies in comparison with star forming galaxies \citep{ashall2016,pruzhinskaya2020,briday2022}. The considerable variation of the maximum brightness and decline rate indicates the significant variation of the energy and $^{56}$Ni production in SNe Ia explosions, this fact can lead to a conclusion about the existence of a dispersion of masses of exploding dwarfs \citep{blondin2017}, at least a part of them can have sub-Chandrasekhar mass \citep{childress2015,foley2020}. This conclusion raises the question about the origin of explosions of sub-Chadrasekhar mass to explain the variation of the $^{56}$Ni production during SNe Ia explosions from $\sim 0.1 M_\odot$ to $\sim 1 M_\odot$ \citep{sharon2020,dutta2022}, probably up to $1.23\pm 0.14 M_\odot$ with the full mass of the explosion products $\approx 1.75 \pm 0.25 M_\odot$ \citep{dimitriadis2022} and similar values \citep{silverman2011}.

The chemical composition of WDs also can play a significant role in the diversity of characteristics of SNe Ia explosions. The most massive WDs are ONe WDs due to the carbon burning in their progenitors \citep{iben1986a}, the mass of such dwarfs is greater than the solar mass \citep{hachisu2019}. Although CO WDs also can be massive (up to the Chandrasekhar limit) both due to the dependence between the initial and remnant masses of the single star (that can be a component of a binary where the mass of the progenitor of the WD and the mass of the WD can additionally grow) and due to the merger of WDs of smaller masses \citep{bogomazov2009,althaus2021,wu2022}, the mass of a CO WD can drop to $0.6 M_\odot$. The existence of ONe dwarfs is reliably confirmed using the analysis of the chemical composition of envelopes of novae, up to one third of dwarfs in them are ONe WDs \citep{gilpons2003}.

The modelling of explosions of sub-Chandrasekhar ONe dwarfs showed their similarity to explosions of CO dwarfs \citep{marquardt2015}. Masses of exploding ONe dwarfs are $1-1.4 M_\odot$ \citep{arnett1969,fenn2016}. Explosions of ONe WDs potentially lead to the formation of low mass ($\approx 1 M_\odot$) neutron stars \citep{shen2012,gvaramadze2019,liud2020,guo2021}. The energy of nuclear burning into Fe of the ONe matter is about 1.8 times less than the energy of the CO matter burning \citep{timmes2000}. Therefore explosions of CO WDs can form about 0.5 mag brighter supernovae in comparison with explosions of ONe WDs for equal masses CO and ONe WDs. This difference potentially is able to explain at least a part of observed dispersion of the maximum brightness of SNe Ia. In particular, such explanation of the offsets in SNe Ia brightnesses in passive and star-forming galaxies mentioned above can be as follows. Star forming galaxies contain the full variety of exploding WDs (ONe, CO) with almost the full range of possible masses of WDs (may be, except the lowest masses, because they require a long period of time to form). Therefore, the brightest SNe Ia from mergers of most massive CO WDs can be found in star-forming galaxies. In passive galaxies the average mass of CO WD mergers should be less than in star-forming galaxies \citep{bogomazov2009}. These relatively low masses of CO WD mergers along with the reduced energetics of ONe WD explosions can lead to the low brightness of SNe Ia in passive galaxies in comparison with star-forming galaxies.

The increasing depth of SNe Ia studies shows the diversity of this subclass of supernovae. In a lot of theoretical works they had been taken as a standard following the Chandrasekhar mass dwarf explosion. Recent studies of faint subclass of SNe Ia named SNe Iax suggested that they are likely result of the WD shell explosion in a close binary system with a helium companion \citep{zeng2022a,zeng2022b,liu2022,ge2022}. The SN Iax class is difficult to reconcile with only the Chandrasekhar mass WD explosion. A detailed theory of SNe Ia is still a matter to develop, one of possible approaches is the comparison of the SN energy with the mass of a shell explosion \citep{nomoto1982,iben1987,tutukov1992b}. Numerical modelling demonstrated that thermonuclear explosion of a sub-Chandrasekhar mass WD with a massive helium shell is consistent with photometric evolution of SNe Ia \citep{dong2022}. Their chemistry admits a part of them are explosions of sub-Chandrasekhar mass dwarfs \citep{tiwari2022}.

A few currently known examples of extragalactic SNe Iax evidently demonstrate a low ejecta mass (\citealp{lykou2022}, masses of ejected envelopes are in the limits $0.1-0.4 M_\odot$). Kinetic energies $10^{49}-10^{50}$ erg are in a good agreement with the nuclear energies of expelled envelopes. The presence and parameters of evolved compact post eruption remnants indicate shell explosions of WDs. Apart that there is a large uniform sample of 127 SNe Ia light curves with early UV excess in their light curves \citep{yao2019}. The observed excess can be explained by the shock interaction with a main sequence companion of the exploding WD. In the frames of this model a significant part of mentioned sample should have such stars as companions of WDs \citep{burke2022}. Another explanation of the blue excess consists in the interaction of the SN envelope with the disc wind matter \citep{hachisu2003}.

\begin{figure}
\includegraphics[width=\columnwidth]{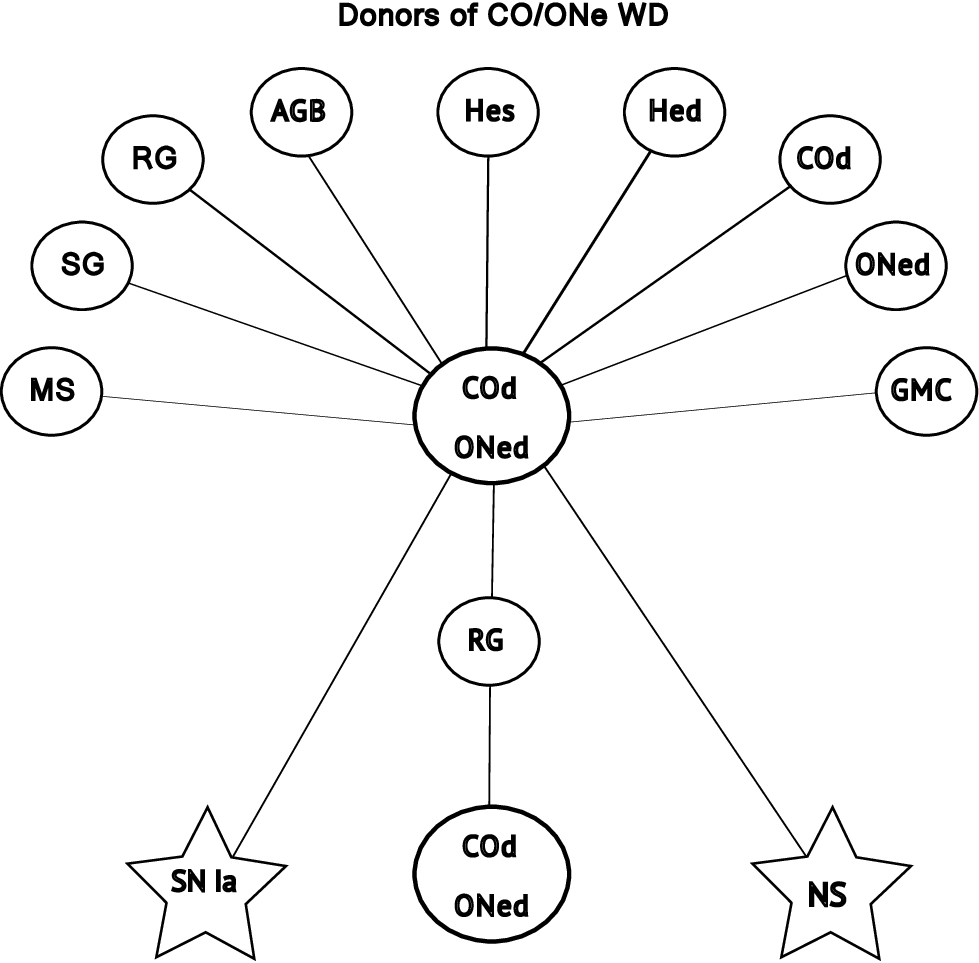}
\vspace{5pt} \caption{Scenarios of explosions of CO WDs (``COd'') and ONe WDs (``ONed''). Here ``MS'' is the main sequence star, ``SG'' is the subgiant, ``RG'' is the red giant, ``AGB'' is the asymptotic giant branch star, ``Hes'' is the non-degenerate helium star, ``Hed'' is the helium ``WD'', ``GMC'' is the giant molecular cloud, ``SN Ia'' is the remnantless SN Ia explosion, ``NS'' is the SN Ia explosion with the formation of the neutron star.}
\label{figure1}
\end{figure}

The main aim of this population synthesis research is to study mergers of double WDs as SNe Ia origin channels. In particular, we calculated dispersions of masses of exploding WDs, the evolution of relative frequencies of ONe SNe Ia (the most massive star in the merger is ONe WD, the second can be ONe, CO, He WD) and CO SNe Ia (the most massive star in the merger is CO WD, the second can be CO, He WD). Dispersions of masses and the evolution relative frequencies can be possible sources of errors in the SNe Ia empirical standardization and conclusions base on this standardization. Below in Introduction we made the overview upon a broader set of possible channels of SNe Ia origin.

The family of donors of CO and ONe WDs exploding as SNe Ia is shown in Figure \ref{figure1}. The initial mass of the donor should be high enough to maintain the growth of the dwarf's mass up to Chandrasekhar limit or to provide the increase of its helium degenerate envelope up to the value that starts the double detonation of the nuclear fuel.

\citet{mazurek1973} considered a possibility of the WD with helium core (He WD) explosion in a binary system. He found that the frequency of such supernovae should be low for the assumption of the total disruption of the exploded dwarf to be consistent with the cosmic element abundances. Here we assume that the He WD cannot explode as a supernova.

In cataclysmic systems with masses of donors $\gtrsim 1 M_\odot$ the rate of the mass exchange stimulated by the magnetic stellar wind can reach up to $\approx 10^{-7} M_\odot$ yr$^{-1}$ \citep{iben1984}. With such accretion rate the persistent hydrogen burning in the envelope of accreting dwarf becomes possible \citep{masevich1988,webbink2002}. The WD with mass $\gtrsim 1 M_\odot$ can accumulate helium envelope with a critical mass that can lead to the double detonation SN Ia. Cataclysmic variables with masses of components $\gtrsim 1 M_\odot$  are potential sources of SNe Ia (Figure \ref{figure1}, ``MS'').

The required accretion rate for the formation of the helium layer with the critical mass can be provided also by red giant donors with masses $\gtrsim 1 M_\odot$ (Figure \ref{figure1}, ``SG'') filling their Roche lobes in pairs with WDs. The mass of the degenerate helium core of the donor should  be $\lesssim 0.5 M_\odot$ \citep{masevich1988}. The donor's mass in this case should be less than the accretor's mass.

The red giant or asymptotic giant branch donor in a symbiotic binary that is close to the filling of its Roche lobe can lose its material at a rate up to $10^{-7}-10^{-5} M_\odot$ yr$^{-1}$ (\citealp{tutukov1972,whelan1973,iben1984b,li2023}, Figure \ref{figure1}, ``RG'', ``AGB''). A substantial part of this matter should be captured by the dwarf's gravity, it can maintain the accretion onto the dwarf and the hydrogen burning in its envelope. As the result the helium layer can be accumulated and a critical mass for the double detonation SN Ia caused by this layer can be reached \citep{devalborro2017}.

An analysis of the evolution of close binary stars demonstrates that the donor in a cataclysmic system can be a non-degenerate helium star (\citealp{iben1987,tutukov1989,wong2021}, Figure 1, ``Hes''). The exchange rate of material between components can reach up to $\sim 10^{-7} M_\odot$ yr$^{-1}$, it can provide the accumulation of the helium envelope with mass enough for the double detonation and the SN Ia explosion \citep{wang2009,wong2021}. Characteristic masses of helium donors in such systems are $0.35-1 M_\odot$ \citep{iben1985,yungelson2005}, their initial masses are $\gtrsim 2.8 M_\odot$ \citep{masevich1988}, the minimum accretor mass is $\approx 0.9 M_\odot$ \citep{wang2009}.

The next version of the evolutionary scenario of the double detonation in Figure \ref{figure1} admits a Roche lobe filling He WD with mass $0.13-0.5 M_\odot$ (``Hed'' in Figure \ref{figure1}). The He WD before the merger should be heated by tidal forces up to high temperatures that should give it the luminosity $\approx 0.15 L_\odot$ \citep{iben1998}. The approach of components under the influence of gravitational waves radiation with subsequent decay of the He WD leads to the origin of a single CO or ONe WD surrounded by the compact helium disc. The accretion of disc's helium can lead to the accumulation of the critical mass by the helium mantle for the dwarf's mass $M_\textrm{WD}\gtrsim 1 M_\odot$. The evolution of WDs with helium donors was discussed, e.g., by \citet{paczynski1967,deloye2005}.

The currently main scenario of SN Ia origin is the merger of a CO or ONe WD with another CO or ONe WD under the influence of the gravitational waves radiation in close binaries (with major semi-axises $\lesssim 3 R_\odot$, Figure \ref{figure1}, ``COd'', ``ONed''). Explosions in this case can be both sub-Chandrasekhar and super-Chandrasekhar \citep{dimitriadis2022,ferrand2022}. Most of existing massive single WDs should be results of mergers of less massive WDs \citep{bogomazov2009,fleury2022,fleury2023,kilic2023,yao2023}. This fact gives more confidence that the merger of WDs in the origin of SNe Ia can play very important role. \citet{dimitriadis2023} argues that the early flux excess combined with late-time oxygen emission of the SN 2021zny supernova indicate the WD merger event. A large amount of observational material on compact detached WD components was accumulated \citep{debes2015,kilic2017,kilic2018,korol2022a,korol2022b}. In the plane $\log P_\textrm{orb}-M$ these systems fill the region $0.01<P_\textrm{orb}/\textrm{d}<10$, $0.25\leq M/M_\odot\leq 1.4$. Several tens of systems have orbital periods $<0.5$ d that lead them to the merger during the time interval less than the Hubble time.

Explosions of SNe Ia also can originate in triple systems, where CO and ONe WDs get closer by the influence of expanded envelope of a third star \citep{iben1999,hamers2013} or can directly collide due to the dynamical interaction \citep{thompson2011,kushnir2013,dong2015}. The frequency of head-on collisions accompanied by SNe Ia explosions should be at least an order of magnitude less than frequency of SNe Ia \citep{hallakoun2019}. \citet{hamers2013} also found the low frequency of SNe Ia originating due to the influence of the third companion.

A new version (it is currently purely hypothetical and can be a matter of future investigations) of the SN Ia origin in Figure \ref{figure1} (``GMC'') is as follows. A CO or ONe WD can get into a giant molecular could. If the density of the cloud is $\rho\approx 10^{-20}$ g cm$^{-3}$, the mass of the dwarf is $M=1 M_\odot$, the size of the cloud is $h\approx 300$ pc then during the travel of the dwarf across the cloud it can accrete the gas of the cloud with mass $\rho V=\rho\pi R^2_\textrm{g} h\approx 5 M_\odot/v_\textrm{s}^4$, where $v_\textrm{s}$ is the velocity of the dwarf with respect to the cloud in units of km s$^{-1}$, $R_\textrm{g}=\dfrac{GM}{v_\textrm{s}^2}$ is the radius of gravitational capture, $G$ is the gravitational constant. Under certain conditions this accreted matter can be enough for the hydrogen burning in the envelope of the dwarf and for the accumulation of the helium with mass required for the double detonation SN Ia. Other versions of SN Ia explosions (considered above) also can take place in a dense molecular cloud. Observational manifestations of such explosions are unknown due to their rarity and a huge optical depth of clouds. Potentially such supernova explosion in a cloud can show itself as a flash in the microwave spectral range with duration about several years depending on the amount of evaporated dust. The temperature of the evaporation of dust determines the size of the dust free zone, the wavelength of radiation (infrared or microwave), the characteristic timescale of the flash.

The helium burning in the mantle of CO or ONe WD instead of the double detonation of nuclear fuel can lead to the origin of a red giant after the expansion of the helium envelope or to the origin of an R CrB type star \citep{masevich1988}. Such short living ($\approx 10^5$ yr) stars actively loosing matter of their envelopes finally leave WDs with initial chemical components.

As was pointed out above, this research is mostly about the double WD merger scenario, its parameters and uncertainties. Nevertheless, it should be kept in mind that the full picture of the SN Ia phenomenon is more complex.

\section{\ttfamily SCENARIO MACHINE}
\label{sec:scm}

{\ttfamily SCENARIO MACHINE} is a specific computer program \citep{kornilov1983} for population synthesis studies of the evolution of close binary stars (see, e.g., a review by \citealp{popov2007}). It is based on the Monte Carlo numerical method. The scientific foundation of the program was described in detail in papers by \citet{lipunov1996,lipunov2009}. To avoid a huge extensive description of the code here we mention only parameters that are important for this study, which often can be used as free parameters.

We assume the power law distribution of initial masses of primary (more massive) stars $f(M_1)\sim M_1^{-\alpha_\textrm{M}}$ ($0.8M_\odot\le M_1\le 120 M_\odot$), the coefficient $\alpha_\textrm{M}$ has two values: $\alpha_\textrm{M}=-2.35$ (the Salpeter mass function, \citealp{salpeter1955,kroupa2019}, usually used in the present study) and $\alpha_\textrm{M}=-2$. The last value is a probable mass function in a low metallic environment in the early universe \citep{tutukov2020} noted as model ``Model P2'', it can be essential due to a potentially very great delay between the formation of binary WDs and their subsequent mergers. Also we assume the flat distribution of the initial separations of components of close binaries in the logarithmic scale \citep{kraicheva1981,abt1983} in the range from $10R_\odot$ to $10^6 R_\odot$, the equiprobable mass ratio of stars in binaries ($f(q)\sim q^0$, where $q=M_2/M_1$, $M_2<M_1$ are the initial masses of components), see, e.g., Equations (1)-(3) in the paper by \citet{lipunov2009}.

The common envelope stage \citep{paczynski1976} of the evolution of close binaries plays an important role in the formation of merging white dwarfs. This stage can bring components much closer than the initial separation and can induce their merger under the influence of gravitational wave radiation in a reasonable interval of time. We used the value of the common envelope efficiency $\alpha_\textrm{CE}=0.5$ (see, e.g., Equation 51 in the description by \citealp{lipunov2009}) as the most suitable for calculations with {\ttfamily SCENARIO MACHINE} \citep{lipunov1996-2}. For stars with initial masses $\leq 10M_\odot$ the code uses the scenario A (see Sections 1-6 in the description by \citealp{lipunov2009}) of the evolution with a weak stellar wind (for our calculations we take the coefficient $\alpha_\textrm{w}=0.3$, Equation 28 by \citealp{lipunov2009}). The stellar wind can play an important role in massive stars, because it changes the mass of the star and the separation between components of the binary. The weak wind in the scenario A leads to the negligible influence of the wind on the main parameters of the evolution of the system (e.g., for the Sun the mass loss rate is $\sim 10^{-13}M_\odot$ yr$^{-1}$, \citealp{wood2002}).

The main sequence evolution of a star with initial mass $\lesssim 0.8 M_\odot$ takes greater time interval than the age of the Universe, other stars finish their lives in a form of WDs, neutron stars and black holes. The minimum initial mass of the star that finally explodes as a supernova and leaves a neutron star or a black hole is $\approx 10 M_\odot$ (see, e.g., \citealp{tutukov1973,ibeling2013}).

\begin{figure}
\includegraphics[width=\columnwidth]{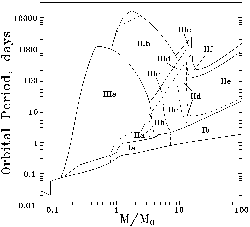}
\vspace{5pt} \caption{The mass transfer diagram \citep{lipunov1996}, see Section \ref{sec:scm} more details.}
\label{diagram}
\end{figure}

The chemical composition of a WD in {\ttfamily SCENARIO MACHINE} code is determined by modes of the mass transfer by \citet{webbink1979} and by the Roche lobe fit by \citet{eggleton1983}; see also \citep{masevich1988}, Figure 47. Following equations are used for this purpose:

$$q_\textrm{p}=\frac{M^\textrm{c}_\textrm{p}}{M^\textrm{s}_\textrm{p}};$$
 
$$f_\textrm{p} = \frac{0.49q_\textrm{p}^{2/3}}{0.6q_\textrm{p}^{2/3}+\ln(1.0+q_\textrm{p}^{1/3})};$$
 
$$f_1 = \frac{0.49}{0.6+\ln 2};$$
 
\begin{align*}
y=-0.9342+1.5\log(a_\textrm{p}(1-e_\textrm{p}))-
0.5\log(M^\textrm{c}_\textrm{p}+M^\textrm{s}_\textrm{p})+ \\
+0.5\log(0.5(1+q_\textrm{p}))+ 1.5\log(f_\textrm{p}/f_1);
\end{align*}

\noindent where $q_\textrm{p}$ is the ratio of masses of components in the stage preceding the WD formation, $M^\textrm{c}_\textrm{p}$ is such preceding mass of the companion star, $M^\textrm{s}_\textrm{p}$ is the preceding mass of the WD progenitor, $f_\textrm{p}$ is the size of the Roche lobe of the WD progenitor in units of the separation of the binary, $f_1$ is such size for equal masses of components (see Equation 15 by \citealp{lipunov2009}), $a_\textrm{p}$ is the preceding semi-major axis of system, $e_\textrm{p}$ is the previous eccentricity of the binary, $y$ is the logarithm of the orbital period in days in the mass transfer diagram (Figure \ref{diagram}) .

Types and modes of the mass exchange are designated using terms by \citet{kippenhahn1967,webbink1979}, the code contains following items concerning WD types (in the form of ``if ... else'' operator; ``II'' is a post main-sequence non-degenerate star with the hydrogen envelope that does not fill its Roche lobe, ``III'' is a Roche lobe filling case):

\begin{itemize}

\item IIA: $x<0.5$.

\item IIB: $y\leq 8.1 - 9x$.
      
\item IIC: $y\leq 11.5 -10x$.

\item IIF: $x>1.085$, $y>  0.20 +1.25x + 0.0167/(x-1.085)$.

\item IIE: $x >1.085$, $y > -1.00 +1.25x + 0.0167/(x-1.085)$.

\item IID: all other supergiant cases.

\item IIIA: $y\leq  3.1 - 2.1(x+0.5)^3$.

\item IIIC: $x\leq 0.84$, $y\leq -0.25 +2.85x$.

\item IIID: $x\leq 1.1$, $y\leq-0.25 +2.85x$.

\item IIIB: $y< 5.57 -3.14x$.

\item IIIE: all other Roche lobe filling cases.

\end{itemize}

\noindent where $x=\log M_\textrm{maxst}/M_\odot$, $M_\textrm{maxst}$ is the maximum mass of a non-degenerate star in the course of its evolution.

For all WD progenitors $M_\textrm{maxst}\leq 10 M_\odot$. Chemical compositions in the code are defined in the following way:

\begin{itemize}

\item ONe WD are result of modes IIID, IIIE, IIC, IID.

\item CO WD are results of modes IIIB, IIIC, IIB.

\item He WD are results of modes IIIA, IIA; if the main sequence star directly evolves to a WD the dwarf is also He WD.

\end{itemize}

Generally, helium WDs are remnants of the evolution of stars with initial masses $0.8-3 M_\odot$. In the binary the mass of the He dwarf is determined by the mass of the donor star in the beginning of its Roche lobe filling: $M_\textrm{He}/M_\odot=0.1 (M/M_\odot)^{0.28}$ \citep{iben1986b}. Conditions of the He WD formation are following: the initial mass of the secondary star $\approx 0.8-3 M_\odot$, the initial semimajor axis in the system ``CO or ONe dwarf + MS star'' $\lesssim 300 R_\odot$ \citep{masevich1988}. The assumptions in the code are close to these models.

According to the current knowledge, stars with masses $8-10M_\odot$ on the main sequence produce ONe WDs with masses $1.1-1.4M_\odot$ \citep{iben1986b,iben1986a,gil-pons2001,menon2013,ibeling2013,denissenkov2013}. We have to emphasize here that borders of initial masses of stars that produce ONe WDs remain rather uncertain up to now despite of the fact of their existence as components of cataclysmic binaries. Therefore we used two different definitions of ONe WDs, the first arises from the mass transfer diagram in the form described above, in the second ONe WD has strict limits $8-10M_\odot$ on the initial mass of the progenitor.

The code is based on evolutionary models with solar metallicity. The value of metallicity can influence several important evolutionary characteristics: the mass loss rate via the stellar wind, the lifetime of the star in a definite evolutionary stage, the mass of the remnant.

The mass loss due to the stellar wind of relatively low mass stars (with massess $<10 M_\odot$) is low for the main sequence stars (see above) and, therefore, any reasonable variations of this parameter in our models (using $\alpha_\textrm{w}$) cannot significantly change the output of the evolution of studied types of binaries.

The gravitational wave timescale for the binary WD mergers can be as long as the age of the Universe and even longer. Therefore the properties of SNe Ia and their frequencies of explosions can reflect the interstellar environment and conditions of the formation of WD progenitors in a distant past. There are data (\citealp{madau2014}, Figure 14) about metallicity of Ly$\alpha$ damped objects (they are of a galactic scale), the metallicity of them \citep{rafelski2012} is $Z\sim 10^{-3}$ for the redshift 2 and $Z\sim 10^{-4}$ for the redshift 3 (the redshift range of the maximum of star formation rate in the Universe according to Figure 8 by \citealp{madau2014}, see below).

\citet{pietrinferni2013} studied the evolution of stars with masses $1 M_\odot$ and $4 M_\odot$ for the wide range of metallicity $Z$. The hydrogen burning time for the $4 M_\odot$ star (Table 2, \citealp{pietrinferni2013}) with metallicity $Z\approx 10^{-4}$ is $\approx 82$\% of the lifetime for the solar metallicity $\sim 10^{-2}$, the order of magnitude of both values is $\sim 10^8$ yr. This difference can be treated as negligible within the frames of the binary population synthesis method and uncertainties of other parameters in this study. For the $1 M_\odot$ star (Table 1, \citealp{pietrinferni2013}) the hydrogen burning of the star with $Z\sim 10^{-4}$ is $\approx 60$\% of the hydrogen burning time of the star with $Z\sim 10^{-2}$, the magnitude of both values is several billion years. From these considerations we can conclude that the evolution of ONe and CO WDs is adequately taken into account, whereas the results concerning He WDs can be moderately affected by the insufficiently correct timescale (see below the model named ``elliptical galaxy'').

The remnant mass in both cases (Tables 1 and 2 by \citealp{pietrinferni2013}) for $Z\sim 10^{-4}$ is about 7\% greater than for $Z\sim 10^{-2}$. This difference is negligible within the accuracy of the method used in this paper.

\section{Population synthesis}
\label{sec:pop-synth}

\begin{table*}
\centering
\caption{Frequencies of events potentially related with SNe Ia calculated for a ``spiral galaxy'', all stars assumed to be binaries. Notations are following: ``ONe'' is the oxygen-neon white dwarf, ``CO'' is the carbon-oxygen white dwarf, ``He'' is the helium white dwarf, ``+'' is the merger of corresponding white dwarfs, ``SD'' is the single accreting WD that potentially can explode, (a) all WDs, (c) the mass of the more massive WD among two merging WDs (or the mass of the dwarf pair with the non-degenerate donor star) is $\geq M_\textrm{max}$ (sub-Chandrasekhar model), (t) the total mass of merging WDs is greater that the Chandrasekhar mass (super-Chandrasekhar model), (8) the initial mass of the progenitor of ONe WD is $\geq 8 M_\odot$ and $<10 M_\odot$. ``SD included'' shows the merger frequencies that include binaries with accreting dwarfs with non-degenerate companions in a supposition that such dwarfs do not explode and their subsequent evolution produce a merger of two WDs, ``SD excluded'' shows the merger frequencies of WDs in a supposition that accreting white dwarfs with non-degenerate companions explode as SNe Ia and the subsequent evolution of their systems do not produce mergers of WDs. ``Model P2'': $\alpha_\textrm{M}=2$, $M_\textrm{max}=1 M_\odot$.}
\label{tab:frequencies}
\begin{tabular}{@{}lllllll@{}}
\hline
Event & \multicolumn{6}{c}{Frequency, per year} \\
 & \multicolumn{3}{c}{SD included} & \multicolumn{3}{c}{SD excluded} \\
 & \multicolumn{2}{c}{$M_\textrm{max}/M_\odot$, for (c)} & Model & \multicolumn{2}{c}{$M_\textrm{max}/M_\odot$, for (c)} & Model \\
 & 1 & 0.8 & P2 & 1 & 0.8 & P2 \\
\hline
ONe+ONe, (a)      &     \multicolumn{2}{c}{$4.5\times 10^{-4}$} & $9.5\times 10^{-4}$ &  $8.4\times 10^{-5}$  &  $3.4\times 10^{-5}$  &  $1.8\times 10^{-4}$ \\
ONe+ONe, (c)      &     $4.2\times 10^{-4}$  &  $4.5\times 10^{-4}$ & $8.9\times 10^{-4}$ &  $5.6\times 10^{-5}$  &  $3.4\times 10^{-5}$  &  $1.1\times 10^{-4}$ \\
ONe+ONe, (t)      &     \multicolumn{2}{c}{$4.5\times 10^{-4}$} & $9.5\times 10^{-4}$ &  $8.4\times 10^{-5}$  &  $3.4\times 10^{-5}$  &  $1.8\times 10^{-4}$ \\
ONe+ONe, (a,8)    &     \multicolumn{2}{c}{$6.1\times 10^{-5}$} & $1.4\times 10^{-4}$ &  $9.7\times 10^{-6}$  &  $9.7\times 10^{-6}$  &  $2.1\times 10^{-5}$ \\
ONe+ONe, (c,8)    &     $6.1\times 10^{-5}$  &  $6.1\times 10^{-5}$ & $1.4\times 10^{-4}$ &  $9.7\times 10^{-6}$  &  $9.7\times 10^{-6}$  &  $2.1\times 10^{-5}$ \\
ONe+ONe, (t,8)    &     \multicolumn{2}{c}{$6.1\times 10^{-5}$} & $1.4\times 10^{-4}$ &  $9.7\times 10^{-6}$  &  $9.7\times 10^{-6}$  &  $2.1\times 10^{-5}$ \\
ONe+CO, (a)       &     \multicolumn{2}{c}{$2.3\times 10^{-3}$} & $4.9\times 10^{-3}$ &  $1.1\times 10^{-3}$  &  $7.6\times 10^{-5}$  &  $2.4\times 10^{-3}$ \\
ONe+CO, (c)       &     $1.2\times 10^{-3}$  &  $2.3\times 10^{-3}$ & $2.7\times 10^{-3}$ &  $5.7\times 10^{-5}$  &  $6.6\times 10^{-5}$  &  $1.2\times 10^{-4}$ \\
ONe+CO, (t)       &     \multicolumn{2}{c}{$2.3\times 10^{-3}$} & $4.9\times 10^{-3}$ &  $1.2\times 10^{-3}$  &  $7.2\times 10^{-5}$  &  $2.3\times 10^{-3}$ \\
ONe+CO, (a,8)     &     \multicolumn{2}{c}{$7.5\times 10^{-4}$} & $1.7\times 10^{-3}$ &         0             &   0                   &  0                  \\
ONe+CO, (c,8)     &     $7.5\times 10^{-4}$  &  $7.5\times 10^{-4}$ & $1.7\times 10^{-3}$ &         0             &   0                   &  0                  \\
ONe+CO, (t,8)     &     \multicolumn{2}{c}{$7.5\times 10^{-4}$} & $1.7\times 10^{-3}$ &         0             &   0                   &  0                  \\
ONe+He, (a)       &     \multicolumn{2}{c}{$4.1\times 10^{-5}$} & $7.1\times 10^{-5}$ &  $4.0\times 10^{-5}$    &  $4.0\times 10^{-5}$    &  $6.9\times 10^{-5}$ \\
ONe+He, (c)       &     $3.9\times 10^{-5}$  &  $4.1\times 10^{-5}$ & $6.7\times 10^{-5}$ &  $3.8\times 10^{-5}$  &  $4.0\times 10^{-5}$    &  $6.9\times 10^{-5}$ \\
ONe+He, (t)       &     \multicolumn{2}{c}{$4.1\times 10^{-5}$} & $7.1\times 10^{-5}$ &  $4.0\times 10^{-5}$    &  $4.0\times 10^{-5}$    &  $6.9\times 10^{-5}$ \\
ONe+He, (a,8)     &     \multicolumn{2}{c}{$6.2\times 10^{-7}$} & $9.1\times 10^{-7}$ &  0                    &  0                    &  0                  \\
ONe+He, (c,8)     &     $6.2\times 10^{-7}$  &  $6.2\times 10^{-7}$ & $9.1\times 10^{-7}$ &  0                    &  0                    &  0                  \\
ONe+He, (t,8)     &     \multicolumn{2}{c}{$6.2\times 10^{-7}$} & $9.1\times 10^{-7}$ &  0                    &  0                    &  0                  \\
CO+CO, (a)        &     \multicolumn{2}{c}{$4.9\times 10^{-3}$} & $8.2\times 10^{-3}$ &  $4.9\times 10^{-3}$  &  $4.5\times 10^{-3}$  &  $8.2\times 10^{-3}$ \\
CO+CO, (c)        &     $3.2\times 10^{-5}$  &  $1.7\times 10^{-3}$ & $6.1\times 10^{-5}$ &  $2.2\times 10^{-5}$  &  $1.3\times 10^{-3}$  &  $4.2\times 10^{-5}$ \\
CO+CO, (t)        &     \multicolumn{2}{c}{$2.1\times 10^{-3}$} & $3.7\times 10^{-3}$ &  $2.1\times 10^{-3}$  &  $1.7\times 10^{-3}$  &  $3.6\times 10^{-3}$ \\
CO+CO, (a,8)        &     \multicolumn{2}{c}{$6.8\times 10^{-3}$} & $1.2\times 10^{-2}$ &  $6.1\times 10^{-3}$  &  $4.6\times 10^{-3}$  &  $1.1\times 10^{-2}$ \\
CO+CO, (c,8)        &     $8.4\times 10^{-4}$  &  $3.6\times 10^{-3}$ & $1.8\times 10^{-3}$ &  $1.3\times 10^{-4}$  &  $1.4\times 10^{-3}$  &  $2.5\times 10^{-4}$ \\
CO+CO, (t,8)        &     \multicolumn{2}{c}{$4.0\times 10^{-3}$} & $1.1\times 10^{-2}$ &  $3.4\times 10^{-3}$  &  $1.8\times 10^{-3}$  &  $6.1\times 10^{-3}$ \\
CO+He, (a)        &     \multicolumn{2}{c}{$2.9\times 10^{-3}$} & $4.3\times 10^{-3}$ &  $2.9\times 10^{-3}$  &  $2.8\times 10^{-3}$      &  $4.3\times 10^{-3}$ \\
CO+He, (c)        &     $7.1\times 10^{-5}$  &  $1.6\times 10^{-4}$ & $1.4\times 10^{-4}$ &  $2.2\times 10^{-5}$  &  $8.3\times 10^{-5}$  &  $4.7\times 10^{-5}$ \\
CO+He, (t)        &     \multicolumn{2}{c}{$1.5\times 10^{-4}$} & $2.9\times 10^{-4}$ &  $1.0\times 10^{-4}$    &  $8.1\times 10^{-5}$  &  $2.1\times 10^{-4}$ \\
CO+He, (a,8)        &     \multicolumn{2}{c}{$2.9\times 10^{-3}$} & $4.4\times 10^{-3}$ &  $2.9\times 10^{-3}$  &  $2.8\times 10^{-3}$      &  $4.4\times 10^{-3}$ \\
CO+He, (c,8)        &     $1.1\times 10^{-4}$  &  $2.0\times 10^{-4}$ & $2.1\times 10^{-4}$ &  $6.0\times 10^{-5}$  &  $1.2\times 10^{-4}$  &  $1.2\times 10^{-4}$ \\
CO+He, (t,8)        &     \multicolumn{2}{c}{$1.9\times 10^{-4}$} & $3.6\times 10^{-4}$ &  $1.4\times 10^{-4}$    &  $1.2\times 10^{-4}$  &  $2.8\times 10^{-4}$ \\
CO SD, (c)        &     -                    &   -                  & -                   &  $1.9\times 10^{-3}$  &   $7.2\times 10^{-3}$ &  $3.5\times 10^{-3}$ \\
CO SD, (c,8)        &     -                    &   -                  & -                   &  $2.8\times 10^{-3}$  &   $9.5\times 10^{-3}$ &  $5.5\times 10^{-3}$ \\
ONe SD, (c)       &     -                    &   -                  & -                   &  $2.8\times 10^{-3}$  &   $4.2\times 10^{-3}$ &  $6.1\times 10^{-3}$ \\
ONe SD, (c,8)     &     -                    &   -                  & -                   &  $1.9\times 10^{-3}$  &   $1.9\times 10^{-3}$ &  $4.1\times 10^{-3}$ \\
CO SD, (t)        &     -                    &   -                  & -                   &  \multicolumn{2}{c}{$5.7\times 10^{-5}$}      &  $1.2\times 10^{-4}$ \\
CO SD, (t,8)        &     -                    &   -                  & -                   &  \multicolumn{2}{c}{$1.1\times 10^{-4}$}      &  $2.4\times 10^{-4}$ \\
ONe SD, (t)       &     -                    &   -                  & -                   &  \multicolumn{2}{c}{$9.7\times 10^{-5}$}      &  $2.2\times 10^{-4}$ \\
ONe SD, (t,8)     &     -                    &   -                  & -                   &  \multicolumn{2}{c}{$4.6\times 10^{-5}$}      &  $1.0\times 10^{-4}$ \\
\hline
\end{tabular}
\end{table*}

We perform calculations of 45 million evolutionary tracks of binary stars using {\ttfamily SCENARIO MACHINE} (30 million tracks for ``Model P2''). The lower limit of the mass in the Salpeter distribution is taken to be $0.8M_\odot$ (the lowest mass of a main sequence star which lifetime is less or equal to the age of the Universe). Because of the features of Monte Carlo simulations using {\ttfamily SCENARIO MACHINE} a meaningful result for rare cases distributed over a very long time interval potentially can be obtained by calculations of several orders of magnitude greater quantity of evolutionary tracks, such huge extensive computations are almost impossible with the current version of the code. Below we call it ``computational statistics problem'' if these method and code features does not allow to make reasonable conclusions.

We assumed two simple models of star formation: ``spiral galaxy'' and ``elliptical galaxy''. ``Spiral galaxy'' is a model of galaxy with star formation rate equal to the current Milky Way star formation rate. ``Elliptical galaxy'' model depicts an object with mass equal to the mass of our galaxy, all stars in this object born at the same (zero) time (distributions of initial parameters are the same as for a ``spiral galaxy''). The progenitors of all stars in ``elliptical galaxy'' born in conditions that correspond to the state of the Universe at a redshift of the arbitrary ``zero'' point, then the stars evolve and the output reflects the events in such synthetic galaxy from ``zero'' up to now.

We estimate frequencies of events potentially related with SNe Ia in a ``spiral galaxy'' assuming all stars as binaries (Table \ref{tab:frequencies}). To obtain frequencies in a real galaxy these frequencies should be multiplied by the part of binaries among all stars. These events are mergers of WDs with subsequent explosions and potential explosions of accreting WDs in binaries with non-degenerate companions. We decipher following mergers of WDs: ONe with ONe, ONe with CO, ONe with He, CO with CO, CO with He. ONe WDs are taken in two versions mentioned above: chemical composition is defined by the mass transfer diagram, the ONe dwarf is the remnant of the evolution of the non-degenerate star with initial mass $8-10M_\odot$. For the latter case other WDs (with initial masses of progenitors $<8M_\odot$) that are accepted as ONe WDs from the mass transfer diagram are CO WDs (the number of CO WDs grows due to ``former'' ONe WDs). The merger frequencies for mentioned combinations of merging WDs are calculated for WDs with all possible masses, for WDs with the mass of the more massive WD among two merging WDs $> M_\textrm{max}$ ($M_\textrm{max}$ takes values $0.8M_\odot$ and $1 M_\odot$), and for WDs with the total mass of both WDs greater that the Chandrasekhar limit. Also we calculate frequencies of potential explosions of accreting WDs in pairs with non-degenerate companions, for ONe and CO WDs that accumulate matter and their mass grows up to $M_\textrm{max}$ (it also takes values $0.8M_\odot$ and $1 M_\odot$) or up to the Chandrasekhar limit, non-accreting sub-Chandrasekhar dwarfs do not explode. All merger frequencies are calculated for versions ``SD included'' and ``SD excluded''. In ``SD included'' accreting WDs do not explode and in the course of evolution their binary systems can become binary WDs that subsequently merge. In ``SD excluded'' the binary system with the explosion of accreting dwarfs cannot produce the merger of WDs. Frequencies for $\alpha_\textrm{M}=-2$  (``Model P2'') are shown separately, for other frequencies we use $\alpha_\textrm{M}=-2.35$ (Salpeter value).

\begin{figure}
\includegraphics[width=\columnwidth]{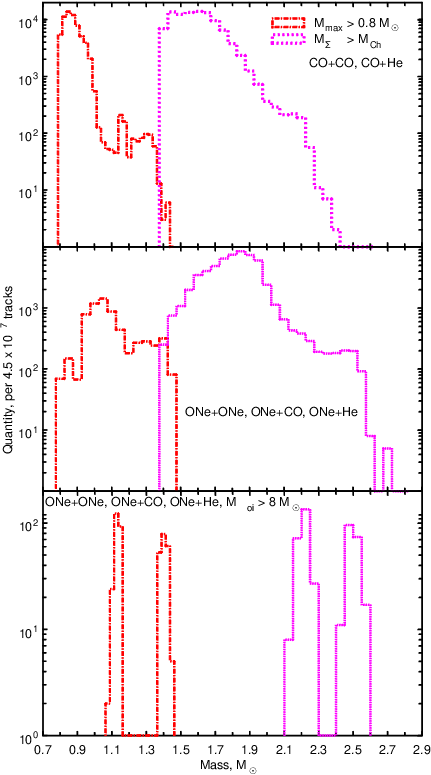}
\vspace{5pt}
\caption{Distributions of merging white dwarfs on the mass, two versions: the mass of the more massive dwarf is $M_\textrm{max}>0.8 M_\odot$, the total mass $M_\Sigma$ of both dwarfs is greater than the Chandrasekhar mass $M_\textrm{Ch}$. Mergers of carbon-oxygen with carbon-oxygen dwarfs (``CO+CO'') and carbon-oxygen with helium dwarfs (``CO+He''). ``SD excluded'' version.}
\label{figure2}
\end{figure}

\begin{figure}
\includegraphics[width=\columnwidth]{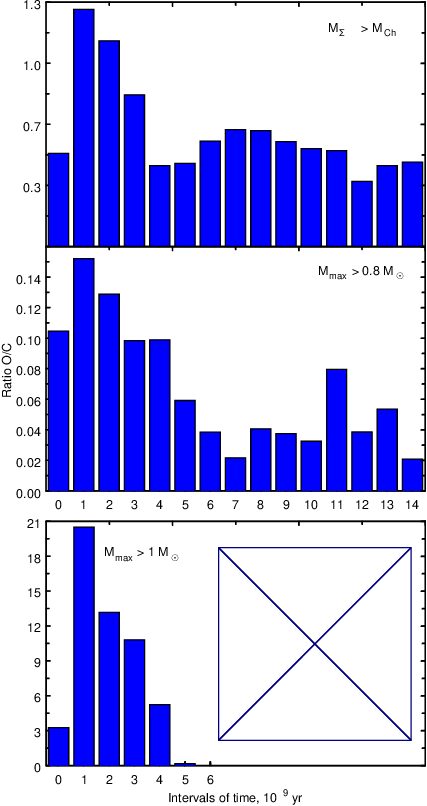}
\vspace{5pt}
\caption{The dependence of the ratio of merger frequencies of ONe dwarfs with other WDs (ONe+ONe, ONe+CO, ONe+He) and CO dwarfs with CO and He WDs (CO+CO, CO+He, except ONe) on time in an ``elliptical galaxy''. ``SD excluded'' version. The crossed out square in the bottom panel is the region that cannot be calculated with {\ttfamily SCENARIO MACHINE} using $M_\textrm{max}=1 M_\odot$.}
\label{figure3}
\end{figure}

Figure \ref{figure2} shows distributions of merging WDs on the mass (``SD excluded'' version), separately for explosions of the more massive WD (with mass $M_\textrm{max}>0.8 M_\odot$) and for super-Chandrasekhar explosions (the total mass $M_\Sigma$ of both WDs is greater than the Chandrasekhar limit $M_\textrm{Ch}$). Distributions are divided into three groups: distributions of mergers with CO WD as the primary component (CO+CO, CO+He), distributions of mergers with ONe WD as the primary component (ONe+ONe, ONe+CO and ONe+He), and the same for ONe mergers for initial masses of progenitors of ONe dwarfs in the range $8-10M_\odot$. According to these distributions merger results have a scatter of masses up to several tens of per cents, both in super- and in sub- Chandrasekhar cases. Strict limits of the ONe initial masses ($8-10M_\odot$) lower the scatter for ONe mergers and show a potential roughness of equations that depict the masses of WD remnants, but even in this case masses of ONe mergers should not be identical.

Figure \ref{figure3} shows the dependence of the ratio of merger frequencies of ONe and CO WDs on time in an ``elliptical galaxy''  (``SD excluded'' version). This ratio is calculated as the quantity of mergers of ONe WDs with other WDs (ONe+ONe, ONe+CO, ONe+He) divided by the quantity of mergers of CO WDs with CO and He WDs (CO+CO, CO+He), the result is averaged over $10^9$ yr. Super- and sub- Chandrasekhar cases are shown separately. The ``computational statistics problem'' does not allow to compute a reliable ratio of frequencies (for the sub-Chandrasekhar model with $M_\textrm{max}> 1 M_\odot$) for the age of the modelled galaxy greater than $\approx 6\times 10^9$ yr (in each 1 billion bin there are only a few mergers with fluctuating numbers of ONe and CO WDs), therefore we show only the first six billion years of the evolution of the $\textrm{O/C}$ ratio. As can be seen from Figure \ref{figure3} the studied ratio can evolve with time. The ``computational statistics problem'' makes it impossible to calculate such ratio for ONe mergers with strict limits on the initial masses of their progenitors ($8-10M_\odot$). The ``SD included'' version increases the quantity of mergers in calculations and makes the presence of the evolution of the frequency ratio less evident, however in our calculations it can fluctuate with a factor of a few.

\section{Discussion and conclusions}

With the help of {\ttfamily SCENARIO MACHINE} population synthesis code we computed frequencies of mergers of WDs with different combinations of their chemical compositions (Table \ref{tab:frequencies}, ONe+ONe, ONe+CO, ONe+He, CO+CO, CO+He; ONe WDs were taken in two versions: (1) the chemical composition is defined by the mass transfer diagram, (2) the initial mass of the progenitor is $8-10M_\odot$). Calculations were performed for all WDs independently on their masses, for sub-Chandrasekhar explosions (using different values of the minimum mass of the more massive WD among two merging dwarfs $M_\textrm{max}$) and for super-Chandrasekhar sum of masses of merging WDs. These operations were made for ``SD included'' and ``SD excluded'' versions assuming non-exploding and exploding accreting WDs in binaries with non-degenerate companions correspondingly. The slope of the initial mass function of stars $\alpha_\textrm{M}$ was varied ($-2.35$ and $-2$).

We have modelled as well distributions of merging WDs on the mass (Figure \ref{figure2}), separately for sub-Chandrasekhar explosions (the mass of the more massive dwarf is $M_\textrm{max}>0.8 M_\odot$) and for super-Chandrasekhar explosions (the total mass $M_\Sigma$ of both dwarfs is greater than the Chandrasekhar mass $M_\textrm{Ch}$). These distributions are shown for CO WDs as primary dwarfs in merging pairs of WDs (CO+CO and CO+He mergers) and for ONe WDs as primaries in merging binaries. Then we calculated the ratio of frequencies of such mergers (CO+CO, CO+He merger frequency to ONe+ONe, ONe+CO, ONe+He merger frequency; Figure \ref{figure2}). ``SD excluded'' version is assumed for figures (accreting dwarfs with non-degenerate companions can explode).

A comparison of merger frequencies in Table \ref{tab:frequencies} provides for several conclusions. The frequency of mergers of CO WDs as primaries is weakly different in ``SD included'' and ``SD excluded'', whereas the frequency of mergers of ONe WDs as primaries is reduced by 5-10 times in ``SD excluded'' model in comparison with ``SD included'' model. These differences reflect the fact that CO WDs usually do not accrete appropriate amount of matter to become more massive than $M_\textrm{max}$ and in our sets of model parameters they do not explode in usually so-called SD (single degenerate) scenario, so they can survive for the DD (double degenerate) scenario. At the same time most of ONe WDs (in the definition from the mass transfer diagram) or even all of them (for initial masses $8-10M_\odot$) are more massive than $M_\textrm{max}$ and in the course of accretion in our model they accumulate enough matter to explode in a binary with a non-degenerate companion. The sub-Chandrasekhar merger frequency of CO dwarfs (CO+CO, CO+He) strongly depends on the $M_\textrm{max}$ (the difference can be up to $\approx 50$ times), because the number of CO dwarfs rapidly grows with the decrease of their masses.

The change of the initial mass function slope (from $-2.35$ to $-2$) leads to a significant growth (up to $\approx 2$ times) of the merger frequency for most of combinations of chemical compositions of merging dwarfs. There are evidences that the slope of the initial mass function of the stars depends on heavy element abundances \citep{shustov2018,tutukov2020}, $\alpha_\textrm{M}$ changes from $-2.35$ for the metallicity $Z\approx0.03$ to about $-2$ for $Z\leq 0.005$ \citep{liang2021}. This fact can additionally influence the evolution of merger frequencies and relative contributions of different mergers with the age.

As can be seen from Figure \ref{figure2} the masses of merging dwarfs (both CO and ONe as primaries) can differ by several tens of percent separately for sub- and super- Chandrasekhar explosions. For the assumption of the full burning of their nuclear fuel in the SNe Ia event corresponding scatter can lead to the scatter of the energy released in the explosion and, therefore, to a potential scatter of peak luminosities of SNe Ia. The theoretical description of such scatter can be even much more complicated, because there are several still poorly understood problems. For example, it is not completely clear how great should be the amount of accreted matter for SD explosions (or SD explosions mostly do not happen, because usually the energy of the accreted matter can be released in novae eruptions). Another unclear problem is the part of the dwarf's mass that undergo nuclear explosion in both SD and especially in DD scenarios, potentially this part can be varied in a very wide range, therefore there is a possibility of a very high scatter of SNe Ia luminosities.

Figure \ref{figure3} shows the evolution of the contribution of ONe and CO mergers in the potential SNe Ia frequencies. As can be seen from this figure, the frequency of super-Chandrasekar ONe (ONe+ONe, ONe+CO, ONe+He) mergers is greater than the frequency of CO (CO+CO, CO+He) mergers in bins 2 and 3 billions after the start of the evolution of an ``elliptical galaxy'', later the frequency of CO mergers becomes greater. In sub-Chandrasekhar mergers with $M_\textrm{max}=0.8 M_\odot$ CO mergers dominates with a factor $\sim 10$. Sub-Chandrasekhar mergers with $M_\textrm{max}=1 M_\odot$ are dominated by ONe mergers during several billions of years, but the ratio of frequencies rapidly drops with time. This demonstrates the extreme importance of studies of ratios of different merger type, their parameters, a potential evolution of relative abundances and real or imitated evolution of SNe Ia brightness. It is necessary to examine a relation between the relative weakness of explosions of ONe WDs and their light curves by numerical models and to find differences (chemistry, light curves, velocities of envelopes, etc.) between explosions of CO and ONe WDs.

A naive example of a cosmologically important conclusion that can be made from the bottom panel of Figure \ref{figure3} is following. Star formation densities in Figure 8 by \citet{madau2014} have maximum in the redshift range 1-3 (infrared estimations give 1-2.5, ultraviolet maximum range is 2-3). We can calculate the time after the maximum of the star formation using simple equation

\begin{equation*}
t(z)=\frac{2}{3H_0(1+z)^{3/2}},
\end{equation*}

\noindent where $z$ is the redshift, $H_0$ is the Hubble constant at $z=0$, $t(z)$ is the time elapsed from the Big Bang at redshift $z$. The time-redshift calculations give (for $H_0=62$ km s$^{-1}$ mpc$^{-1}$): $t(z=2)\approx 2\times 10^9$ yr, $t(z=1)\approx 3.7\times 10^9$ yr, $t(z=0.5)\approx 5.8\times 10^9$ yr. If the evolution in our ``elliptical galaxy'' starts at $z=2$ then in the range 1-2 billion years of evolution ONe mergers start their domination (at $z=1$), several billion years later ($z=0.5$) CO merger frequencies becomes comparable to ONe frequency and, maybe, CO mergers start to dominate. According to lower panel of Figure 4 and 5 by \citep{riess1998}, SNe Ia are dimmer than expected in the redshift range $0.5-1$, this difference is usually explained by the accelerated expansion of the Universe. But Figure \ref{figure3} demonstrated another potential explanation of dim SNe Ia at $z=0.5-1$. Closer to us than $z=0.5$ we see mostly CO mergers and bright SNe Ia, distant SNe Ia are dominated by ONe mergers and dim SNe Ia with a relatively low energy \citep{timmes2000} in the decelerating Universe. SNe Ia in stellar populations with ONe WDs explosions are weaker by $\approx 0.5$ mag that is comparable with the appropriate ``cosmological'' dimming in the accelerated Universe. ``Decelerated'' distant SNe Ia \citep{riess2004,jones2013} are bright CO explosions that can be seen easier than distant faint ONe explosions due to trivial selection effects. All SNe Ia are ``decelerated''. Of course, this conclusion severely depends on assumed model parameters, the middle panel of Figure \ref{figure3} practically excludes this explanation due to overwelmingly greater frequency of CO explosions in comparison with ONe explosions.

Summarizing we can indicate following important problems that should be carefully studied from theoretical point of view:

\begin{itemize}

\item Accurate models of explosions of CO and ONe WDs as SNe Ia for different masses of dwarfs and their exploding shells.

\item Reliable conditions of the formation of ONe WDs.

\item The final chemistry of explosions of CO and ONe WDs.

\item A potential presence of subfamilies of observed SNe Ia explosions (which can have specific dependencies ``light curve shape --- maximum luminosity'' for each separate type), a solution of problems of standardization of SNe Ia for cosmology.

\item The evolution of (from the earliest stars in the Universe up to now): (i) the shape of the initial mass function of binary stars, (ii) the star formation rate, (iii) the metallicity in star forming regions.

\end{itemize}

The observational calibration and usage as a standardized candle of SNe Ia potentially can be biased by the fact that SNe Ia can be results of different channels which relative contribution to birth frequency and observational parameters of SNe Ia can evolve with time. Even in the population with the same age parameters of SNe Ia progenitors can have significant scatters of parameters. There is a possibility that SNe Ia with the same shape of the light curve can have intrinsically different properties \citep{ashall2016}. Also SNe Ia splitted \citep{zhang2020} to ``fast'' and ``slow'' (in Si II 6355) do not show a reduction in Hubble residuals \citep{pan2022}. Theoretical studies should help observational efforts (e.g., \citealp{wiseman2023}). These problems require the development of more detailed observational classification of SNe Ia with theoretical explanations of features of different channels of SNe Ia origin, especially because the output of standardization process can be strongly dependent on the age of the stellar population \citep{lee2022}. Appropriate attempts had been carried out (e.g., \citealp{blinnikov2006,woosley2007,ruiter2013}), but they do not cover all theoretical models of SNe Ia production.

Mergers of WDs should make at least a substantial part of SNe Ia or even be the primary source of this SNe type \citep{tutukov1994,jorgensen1997,panchenko1999,tutukov2002,lipunov2011,yungelson2017}, therefore questions raised in the present study have high importance. Of course, our study does not cancel other models of the SNe Ia origin (see Introduction and, e.g., \citealp{ruiter2020}). Considered progenitors of SNe Ia and uncertainties of parameters potentially can influence estimations of the dark energy role in the geometry and evolution of the Universe based on standardised SNe Ia. The most specially tuned sets of values of parameters of the evolution and explosion of WDs and a specially tuned star formation history function potentially can explain the dependence of the magnitude of SNe Ia on the redshift instead of the accelerated expansion of the Universe. Therefore most of mentioned theoretical problems should be solved before a reliable conclusion can be made, see also, e.g., a recent paper by \citet{miller2023}.

\citet{bogomazov2009} demonstrated that the total mass of merging CO WDs should decrease with time from $\approx 2.1 M_\odot$ (at an age $\approx 10^8$ yr) to $\approx 1.6 M_\odot$ (at an age $\approx 10^{10}$ yr). This hypothetical fall of SNe Ia energetics potentially can explain (at least partially) observed relatively low brightness of SNe Ia (by $\approx 0.5$ mag) in old passive galaxies \citep{ashall2016}. This evolution also can affect the standardization process and cosmological conclusions made using SNe Ia, however, the significance of the influence of the total mass of merging WDs in SNe Ia stays unknown, the mentioned fall is difficult to quantitatively estimate.

In addition we would like to note that SNe Ia appearance and cosmological estimations made using them as standardisable candles can be affected not only by the model uncertainties, but also by usually unaccounted gray extinction in host galaxies of SNe Ia \citep{bogomazov2011}. Studies of extinction of SNe Ia seem to be able to significantly reduce the dispersion of their peak luminosities (see, e.g., \citealp{wang2009,wojtak2023}), or only ``pure'' SNe Ia (which are accepted as free from ordinary and additional gray absorption) can be taken into account \citep{pruzhinskaya2011}. But the gray part of the extinction seems to rather exist than not \citep{gontcharov2023}. A research of its possible value and spatial distribution also would be an interesting issue to theoretical studies along with the model parameters, however, it is out of the scope of the present study.
 
\section*{Acknowledgements}

We are grateful for anonymous referees for the interest to our results and for useful comments that helped to significantly improve the clarity of presentation. The work was supported by the Scientific and Educational School of M. V. Lomonosov Moscow State University ``Fundamental and applied space research'' and by the Russian Science Foundation grant 	23-12-00092 (AIB). AIB acknowledges support from the M. V. Lomonosov Moscow State University Program of Development. AVT is grateful to I.~M. Kulikov and S.~E. Woosley for discussions.

\section*{Data availability}

The data underlying this article will be shared on reasonable request to corresponding authors.

\bibliographystyle{mnras}
\bibliography{bogomazov-tutukov-sn-1a}

\bsp	
\label{lastpage}
\end{document}